\documentclass[aps,pra,superscriptaddress,amsmath,amssymb,preprintnumbers,twocolumn,showpacs]{revtex4}

\usepackage{amssymb}
\usepackage{graphicx}
\usepackage{bm}
\usepackage{color}
\usepackage{cancel}
\usepackage{slashed}

\newcommand{\Ignore}[1]{}

\newcommand{\ot}{{\,\otimes\,}}
\newcommand{\Ket}[1]{\left\vert #1\right\rangle}
\newcommand{\Bra}[1]{\left\langle #1\right\vert}

\renewcommand{\eqref}[1]{Eq.~(\ref{#1})}

\def\<{\langle}
\def\>{\rangle}

\def\e{\mathrm{e}}
\def\ii{\mathrm{i}}

\begin{document}

\title{Generalized Interaction-Free  Evolutions}

\author{Benedetto Militello}
\address{Dipartimento di Fisica e Chimica, Universit\`a di Palermo, Via Archirafi 36, I-90123 Palermo, Italy}
\author{Dariusz Chru\'sci\'nski}
\address{Institute of Physics, Faculty of Physics, Astronomy and Informatics Nicolaus Copernicus University, Grudziadzka 5/7, 87�100 Torun, Poland}
\author{Antonino Messina}
\address{Dipartimento di Fisica e Chimica, Universit\`a di Palermo, Via Archirafi 36, I-90123 Palermo, Italy}
\author{Pawe{\l} Nale\.zyty}
\address{Institute of Physics, Faculty of Physics, Astronomy and Informatics Nicolaus Copernicus University, Grudziadzka 5/7, 87�100 Torun, Poland}
\author{Anna Napoli}
\address{Dipartimento di Fisica e Chimica, Universit\`a di Palermo, Via Archirafi 36, I-90123 Palermo, Italy}

\begin{abstract}

A thorough analysis of the evolutions of bipartite systems characterized by the \lq effective absence\rq\, of interaction between the two subsystems is reported.
First, the connection between the concepts underlying Interaction-Free
Evolutions (IFE) and Decoherence-Free Subspaces (DFS) is explored,
showing  intricate relations between these concepts.
Second, starting from this analysis and inspired by a generalization
of DFS already known in the literature, we introduce the notion of generalized IFE (GIFE), also
providing a useful characterization that allows to develop a general scheme for finding GIFE states.

\end{abstract}

\pacs{03.65.Ta,03.65.Aa,42.50.Ct}

\maketitle

\section{Introduction}\label{sec:Introduction}

Quantum systems are intrinsically subject to relaxation and
dephasing phenomena caused by their unavoidable coupling with the
surrounding~\cite{ref:Alicki,ref:Breuer}. A lot of effort has been
made over the last decades in order to protect quantum systems
from the detrimental effects of the interaction with their
environment~\cite{ref:deco1, ref:deco2, ref:deco3, ref:deco4, ref:deco5}. This research area involves basic
concepts of quantum dynamics of realistic
systems~\cite{ref:Schlosshauer,Joos} but undoubtedly the great deal of
attention dedicated to such issues may be traced back to the more
and more growing interest toward the implementation of reliable
nanodevices where the miniaturization obliges to investigate on
their performance treating them as open quantum systems.
It is well known indeed that simple quantum systems can be ideal
candidates to speed up and improve computational operations \cite{QIT}.
However, if it is true that  solving problems with the use of
quantum algorithms is a revolutionary change in the theory of
computational complexity, on the other hand one has to deal with
the fact that decoherence poses a serious obstacle causing
information loss from the system to its environment. Thus the
possibility of having different ways to bypass detrimental effects
due to decoherence, or, generally speaking, the capability of
systematically envisaging states which preserve coherence
properties,   is an appealing research topic.  In this context,
\emph{subradiant}~\cite{ref:subradiant1,ref:subradiant2,ref:subradiant3,ref:subradiant4,ref:subradiant5,ref:subradiant6,ref:subradiant7}
as well \emph{decoherence-free} (DF)
states~\cite{ref:zanardi1,ref:zanardi2,ref:Nalezyty} have gained wide attention
leading to unitary system dynamics. Several papers indeed have
appeared in the last twenty years concerning the preparation of
such states immune from decoherence in different physical contexts
\cite{ref:immune1,ref:immune2,ref:immune3,ref:immune4}. At the same time, starting from the idea that
decoherence can be avoided remaining inside special subspaces that
are protected from the interaction with the environment,  the
theory of DF subspaces and subsystems has been developed, see for
example the review of Lidar and Whiley and references therein
\cite{ref:Lidar,Lidar-book,ref:LidarAdvances}.

Very recently a new class of states for a closed system, namely
\emph{interaction-free evolving} (IFE)  states, has been
introduced~\cite{ref:IFE} also in the cases wherein the system is
governed by a time-dependent Hamiltonian~\cite{ref:IFE-TD}. By
definition an IFE state of a composed system $\mathrm{A}+\mathrm{B}$ is a state that
evolves as if the interaction between the two parts $\mathrm{A}$ and $\mathrm{B}$
were absent thus implying a unitary evolution for both systems $\mathrm{A}$
and $\mathrm{B}$. As pointed out in refs.~\cite{ref:IFE,ref:IFE-TD} the
concept of IFE state is somehow related to that of
decoherence-free state even if the two concepts are still
different in many aspects.

The aim of this paper is to explore in depth the class of dynamics of a compound system characterized by the fact that the interaction between the two subsystems is seemingly not effective. On the one hand, this analysis leads us to an in-depth study of the connection between the already known interaction-free evolutions and the already known decoherence-free evolutions.  On the other hand, and more important, inspired by the notion of generalized DFS, we are brought to the definition of a new and extended class of IFE, that we call Generalized Interaction-Free Evolution (GIFE). More than this, we provide a characterization of the new class of evolutions (which, of course, contains the previously known IFE) in terms of conservation of some functionals, in the sense that a quantum state is a GIFE state if and only if during its evolution some functionals (we will clarify which ones) maintain their initial values. On this basis, we are also in a condition to formulate a recipe
 that gives the possibility of finding the GIFE states for a given Hamiltonian.

The paper is structured as follows. In the next section we first recall the definitions of IFE and DFS, and then start the discussion about the connection of the relevant two concepts. In Sec.~\ref{sec:GeneralizedIFE} we introduce the notion of generalized IFE. In  the subsequent two sections we try to characterize such new class of evolutions. In particular, in Sec.~\ref{sec:AClassOfHam} we describe a class of Hamiltonian operators that admit GIFE states, while in Sec.~\ref{sec:Characterization} we prove some general properties of GIFE states, in particular the fact that during their evolution some functionals (for example any measure of entanglement) are conserved and on this basis we provide a recipe to find, in principle, all GIFE states for any given Hamiltonian. Finally, in Sec.~\ref{sec:Discussion} we summarize the results of this paper and give some conclusive remarks.

\section{IFE vs DFS --- Hamiltonian formulation}\label{sec:ife_vs_dfs}

In this section we analyze the connection between IFE and DFS. To this end, let us first of all recall the two relevant definitions.

{\it IFE --- } Consider a system whose dynamics is governed by an
Hamiltonian which is the sum of an unperturbed term $H_0$ and an
interaction term $H_\mathrm{I}$: $H=H_\mathrm{0}+H_\mathrm{I}$
(we start by considering time-independent operators). We
will say that a state $\Ket{\chi}$ undergoes an Interaction-Free
Evolution if its evolution is essentially governed by the
unperturbed Hamiltonian $H_\mathrm{0}$ up to a phase factor:
\begin{subequations}
\begin{equation}\label{eq:IFE-Def}
  U(t)\Ket{\chi} = \e^{-\ii
  a t}U_\mathrm{0}(t)\Ket{\chi}\,,
\end{equation}
where
\begin{eqnarray}
  U(t) = e^{-\ii (H_\mathrm{0}+H_\mathrm{I})t} \,, \ \
  U_\mathrm{0}(t) =  e^{-\ii H_\mathrm{0} t} \,.
\end{eqnarray}
and `${a}$' is a real number.
\end{subequations}
In particular, if we consider a composite system \lq\lq system (S) +
environment (E)\rq\rq\,  living in $\mathcal{H}_\mathrm{S} \ot \mathcal{H}_\mathrm{E}$
governed by the time-independent Hamiltonian
\begin{equation} \label{eq:GeneralStructure}
  H = H_\mathrm{S} \ot \mathbb{I}_\mathrm{E}  + \mathbb{I}_\mathrm{S} \ot H_\mathrm{E} + H_\mathrm{I} = H_0 + H_\mathrm{I} \ ,
\end{equation}
then $\Ket{\chi}$ is IFE if
\begin{equation}
  U(t) \Ket{\chi} =  e^{-\ii at}\, e^{-\ii H_\mathrm{S}t} e^{-\ii H_\mathrm{E}t} \Ket{\chi} \, ,
\end{equation}
which, for a product state $\Ket{\psi} \ot \Ket{\phi}$, becomes:
\begin{equation}
  U(t) \Ket{\psi} \ot \Ket{\phi} =  e^{-\ii at}\, e^{-\ii H_\mathrm{S}t}\Ket{\psi} \ot e^{-\ii H_\mathrm{E}t} \Ket{\phi} \, .
\end{equation}

{\it DFS --- }  A decoherence-free subspace of a system
($\mathrm{S}$) interacting with its environment ($\mathrm{E}$) is
a subspace $\mathcal{C}_{\rm DFS} \subset \mathcal{H}_\mathrm{S}$
such that the reduced dynamics
\begin{equation}\label{}
  \rho_\mathrm{S}(t) = {\rm tr}[ U(t) \rho_\mathrm{S} \ot \rho_\mathrm{E} U^\dagger(t) ] = U_\mathrm{S}(t) \rho_\mathrm{S}  U_\mathrm{S}^\dagger(t) ,
\end{equation}
for any $\rho_\mathrm{S}$ supported on $\mathcal{C}_{\rm DFS}$ (i.e.
$\rho_\mathrm{S} = \sum_k p_k |\psi_k\>\<\psi_k|$ and $|\psi_k\> \in
\mathcal{C}_{\rm DFS}$). Note, that the evolution of $\rho_\mathrm{S}$ does
not depend upon the initial state $\rho_\mathrm{E}$ of the environment.
It means that for any $\Ket{\psi} \in \mathcal{C}_{\rm DFS}$ and
arbitrary $\Ket{\phi} \in \mathcal{H}_\mathrm{E}$ one has
\begin{equation}\label{eq:SplitDynamics}
  U(t) \Ket{\psi} \ot \Ket{\phi} =  \exp(-iH_\mathrm{S} t )\Ket{\psi} \ot \exp(-iH^\mathrm{eff}_\mathrm{E} t) \Ket{\phi} \ ,
\end{equation}
where $H^{\rm eff}_\mathrm{E}$  denotes an effective environment
Hamiltonian. Now, if in addition one has $H_\mathrm{I} = \sum_\alpha S_\alpha
\ot E_\alpha$, then necessarily $\sum_\alpha S_\alpha \ot E_\alpha
\Ket{\psi} \ot \Ket{\phi} = \Ket{\psi} \ot \sum_\alpha c_\alpha
E_\alpha \Ket{\phi}$  (in Ref. \cite{ref:zanardi} it is proven that
necessary and sufficient condition to have a DFS is that 1) for all
the states in the DFS it is
$S_\alpha\Ket{\psi}=c_\alpha\Ket{\psi}$, and 2) DFS is invariant under the action of $H_\mathrm{S}$. These conditions imply the
previous condition (see also \cite{ref:Lidar,ref:LidarAdvances})) and hence
\begin{equation}\label{}
  H^{\rm eff}_\mathrm{E} = H_\mathrm{E} + \sum_\alpha c_\alpha E_\alpha\ .
\end{equation}
It is, therefore clear that if $\mathcal{C}_\mathrm{E} \subset \mathcal{H}_\mathrm{E}$ satisfies
\begin{equation}\label{}
  e^{-i H^{\rm eff}_\mathrm{E}t} \Big|_{\mathcal{C}_\mathrm{E}} =  e^{-ia t} e^{-i H_\mathrm{E}} \Big|_{ \mathcal{C}_\mathrm{E} }\ ,
\end{equation}
then $\mathcal{C}_{\rm DFS} \ot \mathcal{C}_\mathrm{E}$ is IFE in
$\mathcal{H}_\mathrm{S} \ot \mathcal{H}_\mathrm{E}$. (In the previous equation we have introduced the notation $O|_\mathcal{C}=O\mathrm{\Pi_\mathcal{C}}$, where $\Pi_\mathcal{C}$ denotes projector onto $\mathcal{C}$). For example it happens when
$\mathcal{C}_\mathrm{E}$ is a common eigenspace of $H^{\rm eff}_\mathrm{E}$ and
$H_\mathrm{E}$. In the very special case where $H_\mathrm{E}$ and
$H_\mathrm{E}^\mathrm{eff}$ differ for a global shift, that is,
$\sum_\alpha c_\alpha E_\alpha = c \mathbb{I}_\mathrm{E}$,  there is a huge
IFE subspace $\mathcal{C}_{\rm DFS}
\otimes\mathcal{H}_\mathrm{E}$. Nevertheless, if the two operators
do not commute, there are states of the environment evolving in a
way which is significantly different from the evolution induced by
$H_\mathrm{E}$. This means that the small system $\mathrm{S}$
evolves as if the interaction with the environment were absent,
but the environment somehow \lq feels\rq\, the presence of the
small system.

On the other hand, if there is a collection of IFE subspaces which
involves all the states of a given subspace of
$\mathcal{H}_\mathrm{S}$ and all possible states of the
environment, this clearly implies that the small system evolves as
if the environment were absent, singling out the presence of a
DFS. Stated another way, if
$\mathcal{C}\otimes\mathcal{H}_{\mathrm{E}}^{(\alpha)}$ is a
collection of IFE subspaces labelled by $\alpha$ and if $\oplus
\mathcal{H}_{\mathrm{E}}^{(\alpha)} = \mathcal{H}_\mathrm{E}$, then
$\mathcal{C}$ is a DFS. But it is evident that this last condition
implies the presence of an effective environment Hamiltonian which
commutes with $H_\mathrm{E}$.

All these facts show in a very clear way that the two concepts of
IFE states and DFS are somehow related and that under some
specific hypotheses each of them implies the other. Nevertheless,
there are a variety of situations, which form the biggest class of
possible situations, wherein one can have IFE states but no DFS
(consider the case of a collection of IFE states whose
environmental parts do not span the whole Hilbert space of the
environment) and vice versa (when there is no common eigenstate of
$H_\mathrm{E}$ and $H_\mathrm{E}^\mathrm{eff}$).

So far, the analysis has been developed in a way that fits well
with time-independent Hamiltonians, but we can make analogous
considerations in a way that fits also when the Hamiltonian is
time-dependent.

To this end, let us analyze the evolution of the
composed system in the interaction picture, that is, let $ \tilde{H}_\mathrm{I}(t) = U_0(t) H_\mathrm{I} U^\dagger_0(t)$,
denotes the interaction Hamiltonian in the interaction picture
with respect to the free evolution governed by $H_\mathrm{0} =
H_\mathrm{S} + H_\mathrm{E}$.

A subspace $\mathcal{C}_\mathrm{DFS}
\subset \mathcal{H}_\mathrm{S}$ is DFS iff
$\tilde{H}_\mathrm{I}(t)|_{\mathcal{C}_{\rm DFS} \ot \mathcal{H}_\mathrm{E}} = \mathrm{\Pi}_\mathrm{DFS} \ot H_\mathrm{E}^{\rm eff}(t)$, where $ \mathrm{\Pi}_\mathrm{DFS}$ is the projector to the subspace $\mathcal{C}_{\rm DFS}$.
On the other hand, a subspace $\mathcal{C} \ot \mathcal{H}_\mathrm{E}^{(\alpha)} \subset \mathcal{H}_\mathrm{S} \ot \mathcal{H}_\mathrm{E}$
is IFE iff
$\tilde{H}_\mathrm{I}(t)|_{ \mathcal{C} \ot \mathcal{H}_\mathrm{E}^{(\alpha)}  } = \alpha(t) \mathrm{\Pi}_{\mathcal{C}} \ot \mathrm{\Pi}_{\mathrm{E},\alpha}$, where $\mathrm{\Pi}_{\mathcal{C}} $ and
$\mathrm{\Pi}_{\mathrm{E},\alpha} $ are the projectors to
$\mathcal{C}$ and $\mathcal{H}_\mathrm{E}^{(\alpha)}$,
respectively. Now, if $\bigoplus_\alpha
\mathcal{H}_\mathrm{E}^{(\alpha)} = \mathcal{H}_{E}$, then
$\mathcal{C}$ is DFS, being $H^{\rm eff}(t) = \bigoplus_\alpha \alpha(t) \mathrm{\Pi}_{\mathrm{E},\alpha}$.
Moreover, if $\mathcal{C} \ot \mathcal{H}_\mathrm{E}$ is IFE then
$\tilde{H}_\mathrm{I}(t)|_{ \mathcal{C} \ot
\mathcal{H}_\mathrm{E}}  = \alpha(t) \mathrm{\Pi}_\mathcal{C} \ot
\mathbb{I}_\mathrm{E}$.

These last two assertions clarify very well the connection between IFE and DFS.

\section{Generalized IFE}\label{sec:GeneralizedIFE}

It is worth noting that Eq.~(\ref{eq:SplitDynamics}) shows that
the presence of a DFS implies that the small system (S) evolves
according to its free Hamiltonian, while the environment evolves
through an effective Hamiltonian, which may commute or not with
the environment free Hamiltonian. It somehow resembles an IFE
evolution, where the two systems do not interact, though the
Hamiltonian of one of the two systems (the environment, in this
case) is not the free one.

As another important fact, we mention that the notion of DFS can
be generalized, according to the analysis in
Ref.~\cite{ref:Lidar}, in the following way. Suppose a quantum
system interacting with its environment is describable by the
Hamiltonian, as in \eqref{eq:GeneralStructure},
then a given subspace of the Hilbert space of the system, say
${\cal C}_\mathrm{GDFS} \subset {\cal H}_\mathrm{S}$, is a
generalized DFS if, whatever the state $\Ket{\psi}$ of the
environment, the system prepared in a state $\Ket{\phi} \in {\cal
C}_\mathrm{GDFS}$ evolves as if it was not interacting with the
environment, even if its dynamics is not governed by $H_\mathrm{S}$
but it is determined by an effective system Hamiltonian
$H^\mathrm{eff}_\mathrm{S} \not= H_\mathrm{S}$.

Both these facts suggest a possible extension of the concept of
IFE. Consider a bipartite system $\mathrm{A}+\mathrm{B}$, whose
dynamics is governed by
\begin{equation}\label{eq:ABI_Hamiltonian}
H= H_\mathrm{A} \ot \mathbb{I}_\mathrm{B}+ \mathbb{I}_\mathrm{A} \ot H_\mathrm{B}+H_\mathrm{I} \, .
\end{equation}
We can define generalized IFE (GIFE) those evolutions where each
of the two subsystems undergoes an evolution \textit{seemingly
independent} from the other subsystem. This means that each of the
two subsystems evolves under the action of an Hamiltonian
$H_\mathrm{k}^\mathrm{eff}(t)$, with
$\mathrm{k}=\mathrm{A},\mathrm{B}$, not necessarily coincident
with $H_\mathrm{k}$. More precisely, a state $\Ket{\chi} \in \mathcal{H}_\mathrm{A} \ot \mathcal{H}_\mathrm{B}$ is a
GIFE state if there exist two operators $H_\mathrm{A}^\mathrm{eff}(t)$ and
$H_\mathrm{B}^\mathrm{eff}(t)$ such that the following set
of equations can be satisfied:
\begin{equation}\label{eq:GIFE-def-unitary}
\left\{
\begin{array}{l}
  U(t)\Ket{\chi} = U_\mathrm{A}^\mathrm{eff}(t)\ot U_\mathrm{B}^\mathrm{eff}(t) \Ket{\chi} \,,\\
  \\
  \ii \dot{U}_\mathrm{k}^\mathrm{eff} \, = H_\mathrm{k}^\mathrm{eff}(t) U_\mathrm{k}^\mathrm{eff}(t) \,, \qquad \mathrm{k}=\mathrm{A},\mathrm{B}\,.
\end{array}
\right.
\end{equation}
Note, that if
\begin{equation}\label{}
  |\chi\> = \sum_\alpha \chi_{\alpha\beta} \, |e_\alpha\> \ot |f_\beta  \>
\end{equation}
with $\{e_\alpha\}$ and $\{f_\beta\}$ being orthonormal basis in $\mathcal{H}_\mathrm{A}$ and $\mathcal{H}_\mathrm{B}$, respectively, then
\begin{equation}\label{}
  |\chi(t)\>  = \sum_\alpha \chi_{\alpha\beta}\, U_\mathrm{A}^\mathrm{eff}(t) |e_\alpha\> \ot U_\mathrm{B}^\mathrm{eff}(t) |f_\beta\> .
\end{equation}
It should be clear  that GIFE states are nothing but IFE states with respect to a suitable
effective interaction Hamiltonian. Following Ref.~\cite{ref:IFE-TD} one finds that $|\chi\>$ defines GIFE state if and only if
\begin{subequations}
\begin{equation}\label{eq:GIFE_condition}
 \breve{H}_\mathrm{I}^\mathrm{eff}(t) \Ket{\chi} = 0\,,
\end{equation}
where
\begin{equation}\label{eq:H_I_Eff}
 \breve{H}_\mathrm{I}^\mathrm{eff}(t) \equiv   \breve{H}(t) - \breve{H}_\mathrm{A}^\mathrm{eff}(t) -  \breve{H}_\mathrm{B}^\mathrm{eff}(t)  \,,
\end{equation}
and the new interaction picture is defined as follows
\begin{equation}
\breve{O}(t) = U_\mathrm{A}^{\mathrm{eff}\dag}(t) \ot
U_\mathrm{B}^{\mathrm{eff}\dag}(t)  \, O \,
U_\mathrm{A}^\mathrm{eff}(t) \ot U_\mathrm{B}^\mathrm{eff}(t)\,.
\end{equation}
\end{subequations}

One could think of replacing condition in
\eqref{eq:GIFE_condition} with the seemingly more general
condition $\breve{H}_\mathrm{I}^\mathrm{eff}(t) \Ket{\chi} =
\alpha(t) \Ket{\chi}$, resembling what we have found in
Ref.~\cite{ref:IFE-TD}. However, since in the case of GIFE we have
to find also the two effective unperturbed Hamiltonian operators,
the constant term $\alpha(t)\mathrm{I}$ can be included in such
operators, which makes the two problems  essentially equivalent.

Of course, the standard case of IFE is included as a special case of GIFE. (We will use the expression \lq proper GIFE\rq\, to talk about GIFE states which are not IFE states.)

{\bf An Example} --- In order to better illustrate the notion of GIFE, we will analyze a specific physical situation where both IFE and GIFE arise.
Consider the multi-spin system interacting with a bosonic field (see for example Ref.\cite{ref:LidarAdvances}). The relevant Hamiltonian is given by:

\begin{subequations}
\begin{equation}
H_\mathrm{S} = \sum_k \Omega_k \sigma_z^{(k)}\,,
\end{equation}
\begin{equation}
H_\mathrm{E} = \sum_j \omega_j a_j^\dag a_j\,,
\end{equation}
\begin{equation}
H_\mathrm{I} =  \left(\sum_k \sigma_z^{(k)}\right) \ot \sum_j g_j
(a_j + a_j^\dag))\,.
\end{equation}
\end{subequations}

Since $\sum_k \sigma_z^{(k)}$ is nothing but the total pseudo-spin
(let us call it $J_z$), and $[H_\mathrm{S}, J_z]=0$, we find that
each eigenspace of $J_z$ is decoherence-free. When $\mathrm{S}$ is prepared
in an eigenstate of $J_z$ with eigenvalue $m$, the environment
evolve according to
\begin{equation}
H_\mathrm{E}^\mathrm{eff} = \sum_k \left[\omega_k a^k_k a_k + m
g_k (a_k + a_k^\dag)\right]\,,
\end{equation}
and we have a GIFE subspace, unless $m=0$, in which case we have
an IFE subspace.

Now it comes the crucial question: {\em how to characterize bipartite Hamiltonians giving rise to GIFE states?}
In the next two sections, we will make some efforts in this direction.

\section{A class of Hamiltonians that admits GIFE}\label{sec:AClassOfHam}

The previous example suggests a structure of Hamiltonians that admit GIFE.

Consider the following time-independent Hamiltonian in $\mathcal{H}_\mathrm{A} \ot \mathcal{H}_\mathrm{B}$
\begin{equation}\label{}
  H = H_\mathrm{A} \ot \mathbb{I}_\mathrm{B} + \mathbb{I}_\mathrm{A} \ot H_\mathrm{B} + \sum_k P_k \ot B_k ,
\end{equation}
where $P_k = |k\>\<k|$ are projectors into the computational basis vectors $|k\>$ in $\mathcal{H}_\mathrm{A}$ and $B_k$ are hermitian operators in $\mathcal{H}_\mathrm{B}$. Now, assuming that $H_\mathrm{A} = \sum_k \epsilon_k P_k$ one finds
\begin{equation}\label{}
  H = \sum_k P_k \ot Z_k ,
\end{equation}
where $Z_k = \epsilon_k \mathbb{I}_\mathrm{B} + H_\mathrm{B} + B_k$. Such Hamiltonian leads to a pure decoherence of the density operator $\rho_\mathrm{A}$ of subsystem A:
\begin{equation}\label{}
  \rho_\mathrm{A}(t) = {\rm tr}_\mathrm{B} (e^{-iHt} \rho_\mathrm{A} \ot \rho_\mathrm{B} e^{iHt}) = \sum_{k,l} c_{kl}(t) P_k \rho_\mathrm{A} P_l ,
\end{equation}
with $c_{kl}(t) = {\rm tr}( e^{-i Z_k t} \rho_\mathrm{B} e^{iZ_lt})$. It is clear that each 1-dim. subspace in $\mathcal{H}_\mathrm{A}$ spanned by $|k\>$ defines DFS.
Note, that $|k\> \ot |\phi_\mathrm{B}\>$, where $|\phi_\mathrm{B}\>$ is an arbitrary vector from $\mathcal{H}_\mathrm{B}$, defines GIFE but not IFE. Indeed, one has
\begin{equation}\label{}
 e^{-iHt} |k\> \ot |\phi_\mathrm{B}\> = e^{-iH_\mathrm{A} t}|k\> \ot e^{-i(H_\mathrm{B} + Z_k)t}|\phi_\mathrm{B}\> .
\end{equation}
It is clear that one may replace $H_\mathrm{B}$ and $B_k$ by time-dependent operators.

The previous Hamiltonian structure gives rise to evolutions which are IFE for one subsystem and GIFE for the other one. In the following we give a more general structure for the Hamiltonians that give rise to GIFE evolutions for both subsystems. Consider what follows.

Let us recall \cite{ref:IFE} that $|\chi\> \in \mathcal{H}_\mathrm{A} \ot \mathcal{H}_\mathrm{B}$ satisfying
\begin{equation}\label{0}
  H_{\rm I} |\chi\> = 0\, ,
\end{equation}
is IFE for the Hamiltonian (\ref{eq:ABI_Hamiltonian}) if and only if
\begin{equation}\label{n}
  H_{\rm I} H_0^n|\chi\> = 0\, , \ \ \ n=1,2,3,\ldots\ ,
\end{equation}
where $H_0 =  H_\mathrm{A} \ot \mathbb{I}_\mathrm{B}+ \mathbb{I}_\mathrm{A} \ot H_\mathrm{B}$ is the ``free" part of $H$.
Note that one can always rewrite the total Hamiltonian performing the following ``corrections" of $H_\mathrm{A}$ and $H_\mathrm{B}$:
\begin{equation}\label{corrections}
  H = (H_\mathrm{A} +\Delta_\mathrm{A}) \ot \mathbb{I}_\mathrm{B}+ \mathbb{I}_\mathrm{A} \ot (H_\mathrm{B}+ \Delta_\mathrm{B}) + H^{\rm eff}_\mathrm{I} \,,
\end{equation}
with
\begin{equation}\label{}
  H^{\rm eff}_\mathrm{I} = H_\mathrm{I} - [\Delta_\mathrm{A} \ot \mathbb{I}_\mathrm{B} + \mathbb{I}_\mathrm{A} \ot \Delta_\mathrm{B}]\, .
\end{equation}
Hence the question of the existence of GIFE states is equivalent to the existence of suitable operators $\Delta_\mathrm{A}$ and $\Delta_\mathrm{B}$ such that
\begin{equation}\label{}
  H_\mathrm{A}^\mathrm{eff}= H_\mathrm{A} + \Delta_\mathrm{A} \ , \ \  H_\mathrm{B}^\mathrm{eff}= H_\mathrm{B} + \Delta_\mathrm{B}\ ,
\end{equation}
satisfies conditions (\ref{eq:GIFE-def-unitary}). Here we propose the following class of Hamiltonians admitting GIFE subspaces: let $S_\mathrm{A}$ and $S_\mathrm{B}$ be linear subspaces in $\mathcal{H}_\mathrm{A}$ and $\mathcal{H}_\mathrm{B}$, respectively. Moreover, let $\Pi_\mathrm{A}$ and $\Pi_\mathrm{B}$ be the corresponding orthogonal projectors, that is,
\begin{equation}\label{}
  S_\mathrm{A} = \Pi_\mathrm{A} \mathcal{H}_\mathrm{A}\ , \ \ \  S_\mathrm{B} = \Pi_\mathrm{B} \mathcal{H}_\mathrm{B}\, .
\end{equation}
We construct a class of bipartite Hamiltonians such that any $|\chi\> \in S_{AB} = S_\mathrm{A} \ot S_\mathrm{B}$ is a GIFE state. Let $H_\mathrm{A}$ and $H_\mathrm{B}$ be Hamiltonians of systems A and B, respectively, such that
\begin{equation}\label{}
  [H_\mathrm{A}, \mathrm{\Pi}_\mathrm{A}]=0 \ , \ \ \ [H_\mathrm{B}, \mathrm{\Pi}_\mathrm{B}]=0\, .
\end{equation}
Let $\Delta_\mathrm{A}$ and $\Delta_\mathrm{B}$ be two ``corrections" satisfying the same commutation relations, i.e.
\begin{equation}\label{}
  [\Delta_\mathrm{A}, \mathrm{\Pi}_\mathrm{A}]=0 \ , \ \ \ [\Delta_\mathrm{B}, \mathrm{\Pi}_\mathrm{B}]=0\, .
\end{equation}
Consider now  the interaction part
\begin{equation}
  H_\mathrm{I} = \Delta_\mathrm{A} \ot \mathrm{\Pi}_\mathrm{B} +
  \mathrm{\Pi}_\mathrm{A} \ot \Delta_\mathrm{B} + \Delta^\perp
\end{equation}
with $\Delta^\perp$ be an arbitrary bipartite operator such that
\begin{eqnarray}
   \Delta^\perp \mathrm{\Pi}_\mathrm{A} \ot \mathrm{\Pi}_\mathrm{B} = \mathrm{\Pi}_\mathrm{A} \ot \mathrm{\Pi}_\mathrm{B} \Delta^\perp =
  0\,.
\end{eqnarray}
It is clear that taking $|\chi\> \in S_{\mathrm{AB}}$ one finds in
general $ H_{\rm I} |\chi\> \neq 0$, and hence condition (\ref{0})
is not satisfied. However, correcting $H_\mathrm{A}$ and
$H_\mathrm{B}$ as in (\ref{corrections}) one finds
\begin{eqnarray}
  H &=& H_\mathrm{A} \ot \mathbb{I}_\mathrm{B}+ \mathbb{I}_\mathrm{A} \ot H_\mathrm{B}+  H_{\rm I} \nonumber \\
   &=&  H^{\rm eff}_\mathrm{A} \ot \mathbb{I}_\mathrm{B}+ \mathbb{I}_\mathrm{A} \ot H^{\rm eff}_\mathrm{B} + H^{\rm eff}_{\rm I}\, ,
\end{eqnarray}
where
\begin{equation}\label{}
  H^{\rm eff}_{\rm I} =  \Delta_\mathrm{A} \ot \mathrm{\Pi}^\perp_\mathrm{B} + \mathrm{\Pi}^\perp_\mathrm{A} \ot \Delta_\mathrm{B} + \Delta^\perp\, ,
\end{equation}
with
\begin{equation}\label{}
  \mathrm{\Pi}^\perp_\mathrm{A} = \mathbb{I}_\mathrm{A} - \Pi_\mathrm{A}\ , \ \ \   \mathrm{\Pi}^\perp_\mathrm{B} = \mathbb{I}_\mathrm{B} - \Pi_\mathrm{B}\, .
\end{equation}
It is, therefore, clear that
\begin{equation}\label{0-1}
  H_{\rm I}^{\rm eff} |\chi\> = 0 \, ,
\end{equation}
for any $|\chi\> \in S_{AB}$. Moreover, one easily checks
\begin{equation}\label{n-1}
  H_{\rm I}^{\rm eff} (H_0^{\rm eff})^n|\chi\> = 0 \ , \ \ \ n=1,2,\ldots\, ,
\end{equation}
where $H_0^{\rm eff} = H^{\rm eff}_\mathrm{A} \ot \mathbb{I}_\mathrm{B}+ \mathbb{I}_\mathrm{A} \ot H^{\rm eff}_\mathrm{B}$.
Hence, conditions (\ref{0-1})--(\ref{n-1}) for new effective Hamiltonians are exactly the same as (\ref{0})--(\ref{n}) for the original $H_0$ and $H_{\rm I}$. This proves that  $|\chi\>$ defines GIFE state.

Interestingly, in the special case when $[H_\mathrm{A},
\Delta_\mathrm{A}]=0$ and $[H_\mathrm{B}, \Delta_\mathrm{B}]=0$,
the subspace $S_{\mathrm{AB}}$ can be decomposed into common
eigenspaces of $H_\mathrm{A}$ and $\Delta_\mathrm{A}$, for the
subsystem $\mathrm{A}$ (call them $\mathcal{C}_\mathrm{A}^\alpha$,
where $\alpha$ is the relevant eigenvalue of $\Delta_\mathrm{A}$),
and into common eigenspaces of $H_\mathrm{B}$ and
$\Delta_\mathrm{B}$, for the subsystem $\mathrm{B}$ (call them
$\mathcal{C}_\mathrm{B}^\beta$, where $\beta$ is the relevant
eigenvalue of $\Delta_\mathrm{B}$):
\begin{equation}\label{}
  S_{\mathrm{AB}} = \bigoplus_{\alpha,\beta} \mathcal{C}_\mathrm{A}^\alpha \ot \mathcal{C}_\mathrm{B}^\beta\, .
\end{equation}
The tensor product $\mathcal{C}_\mathrm{A}^\alpha \ot \mathcal{C}_\mathrm{B}^\beta$
of each two of such common eigenspaces
corresponds to a proper IFE subspace, where the dynamics differs
from the unperturbed one by the phase factor $\exp\left(\int_0^t
[\alpha(s)+\beta(s)]\mathrm{d}s \right)$.

\section{Characterization of GIFE}\label{sec:Characterization}

\newcommand{\trc}{\mathrm{tr}}

In this Section we provide general properties of generalized
interaction-free evolutions and derive a general scheme for
finding GIFE states in principle for any given Hamiltonian.

\subsection{General Properties of GIFE}


Let us first of all briefly discuss the relation between GIFE and
entangled states in $\mathcal{H}_\mathrm{A} \ot
\mathcal{H}_\mathrm{B}$. Let us observe that if $\mathcal{E}$ is a
genuine entanglement measure then for any GIFE state $\Ket{\chi}$
one has $\frac{d}{dt}\mathcal{E}(\Ket{\chi(t)}) = 0$, i.e. every
GIFE state is an entanglement-preserving state. Indeed, GIFE
states evolve as if they were under the action of two local
(effective) Hamiltonians, and then, whatever is the entanglement
measure considered, the amount of entanglement does not change in
the evolution of a GIFE state. In particular, the entropy of
entanglement $S(\trc_\mathrm{A} |\chi(t)\>\<\chi(t)|) =
S(\trc_\mathrm{B} |\chi(t)\>\<\chi(t)|)$ and the linear entropy
$S_L= 1 -
\trc_\mathrm{B}\left(\trc_\mathrm{A}|\chi(t)\>\<\chi(t)|\right)^2$
do not depend on time. Moreover, any function of the eigenvalues
of the two reduced density operators, either
$\rho_\mathrm{B}=\trc_\mathrm{A}\Ket{\chi(t)}\Bra{\chi(t)}$ or
$\rho_\mathrm{A}=\trc_\mathrm{B}\Ket{\chi(t)}\Bra{\chi(t)}$, does
not depend on time.

Now we are ready to provide the necessary and sufficient condition for a state to be a GIFE state.

\textbf{Theorem} { \it %
If $\min\{n_\mathrm{A},n_\mathrm{B}\}=n$ (where
$n_{\mathrm{A}/\mathrm{B}} = {\rm
dim}\mathcal{H}_{\mathrm{A}/\mathrm{B}}$), then the state
$\Ket{\chi}$ is GIFE iff
\begin{equation}\label{eq:Condition_rho_k}
  \frac{d}{dt} \trc_\mathrm{B}\left(\trc_\mathrm{A} |\chi(t)\>\<\chi(t)| \right)^k = 0\ ,
\end{equation}
for $k=1,2,\ldots,n$. }

Proof --- Given the set of eigenvalues $p_1, \ldots, p_n$ of the
reduced density operator
$\rho_\mathrm{B}=\trc_\mathrm{A}{\Ket{\chi}\Bra{\chi}}$ (we here
assume $n_\mathrm{B}=n$, otherwise we use $\rho_\mathrm{A}$), the
set of equations in \eqref{eq:Condition_rho_k} turns out to be
equivalent to:
\begin{equation}\label{eq:Condition_p_l_k}
\sum_{l=1}^n p_l^k(t) = s_k\,, \qquad k=1,...n\,,
\end{equation}
where $s_k$'s are $n$ real positive numbers. By the way, the
condition corresponding to $k=1$ is trivial, being nothing but the
normalization.

From the comments we have done just above it is clear that any
GIFE state satisfies all such equations. We then only need to
prove that if a state satisfies \eqref{eq:Condition_p_l_k} then it
is GIFE. If the set of algebraic equations in
\eqref{eq:Condition_p_l_k} (which is solvable through the use of
Newton-Girard identities) is the same at any time, then it also
admits the same solutions ($p_l$'s) at any time. Now, given the
set $p_l$'s, it is well known that the pure state describing the
total system can be put in the following form:
\begin{equation}
\Ket{\chi(t)} = \sum_{l=1}^n \sqrt{p_l}
\Ket{\phi_l(t)}_\mathrm{A}\otimes \Ket{\psi_l(t)}_\mathrm{B}\,,
\end{equation}
where the coefficients $\sqrt{p_l}$, which are nothing but the
Schmidt coefficients of $|\chi\>$, in this case are
time-independent. Moreover, the states
$\Ket{\phi_l(t)}_\mathrm{A}$ and $\Ket{\psi_l(t)}_\mathrm{B}$,
though time-dependent, are two sets of orthonormal states at every
time. Since it is clear that there are two unitary operators,
$U_\mathrm{A}^\mathrm{eff}(t)$ and $U_\mathrm{B}^\mathrm{eff}(t)$,
that map $\Ket{\phi_k(0)}$ into $\Ket{\phi_k(t)}$  and
$\Ket{\psi_k(0)}$ into $\Ket{\psi_k(t)}$, then we can consider the
state $\Ket{\chi}$ as if it evolves unitarily:  $|\chi(t)\> =
U_\mathrm{A}^\mathrm{eff}(t)\ot U_\mathrm{B}^\mathrm{eff}(t)
|\chi\>$.

\subsection{Recipe to find GIFE states}

By exploiting the previous results we propose a strategy to check whether a given Hamiltonian admits GIFE evolutions.
Let us restrict our analysis to the case of time-independent Hamiltonians. Using
\begin{equation}
H\Ket{\lambda_i} = \lambda_i \Ket{\lambda_i}\,,
\end{equation}
we can write the general solution of the relevant Schr\"odinger problem in the following way:
\begin{equation}
\Ket{\chi(t)} = \sum_i c_i \e^{-\i\lambda_i t} \Ket{\lambda_i}\,.
\end{equation}

Now, since $\trc_\mathrm{A} |\chi(t)\>\<\chi(t)|$ can be cast in the following form,
\begin{equation}
\trc_\mathrm{A} |\chi(t)\>\<\chi(t)| = \sum_{ij} c_i c_j^* \e^{-\i (\lambda_i-\lambda_j) t} \trc_\mathrm{A} \Ket{\lambda_i}\Bra{\lambda_j}\,,
\end{equation}
conditions in \eqref{eq:Condition_rho_k} assume the following form:
\begin{widetext}
\begin{subequations}
\begin{eqnarray}
\label{eq:Cond_Normalization}
&k=1:& \,\,\,  \frac{d}{dt} \sum_{i} |c_i|^2 =0\,,   \\
\label{eq:Cond_Purity}
\nonumber
&k=2:& \,\,\,  -\i \sum_{i_1 j_1 i_2 j_2} c_{i_1} c_{j_1}^*c_{i_2} c_{j_2}^* \e^{-\i (\lambda_{i_1}-\lambda_{j_1}+\lambda_{i_2}-\lambda_{j_2}) t} (\lambda_{i_1}-\lambda_{j_1}+\lambda_{i_2}-\lambda_{j_2}) \trc_\mathrm{B}\left( \trc_\mathrm{A} \Ket{\lambda_{i_1}}\Bra{\lambda_{j_1}} \trc_\mathrm{A} \Ket{\lambda_{i_2}}\Bra{\lambda_{j_2}}\right) = 0\,,   \\
\\
\nonumber
&...& \\
\label{eq:Cond_k=n}
\nonumber
&k=n:& \,\,\,  (-\i)^{n-1}\sum_{i_1 j_1... i_n j_n} \left( \prod_{s=1}^{n} c_{i_\mathrm{S}} \right) \times \left( \prod_{s=1}^{n} c_{j_\mathrm{S}}^* \right) \times \\
\nonumber
&& \,\,\, \e^{-\i (\sum_{s=1}^{n} \lambda_{i_\mathrm{S}}-\sum_{s=1}^{n}\lambda_{j_\mathrm{S}}) t} \left(\sum_{s=1}^{n} \lambda_{i_\mathrm{S}}-\sum_{s=1}^{n}\lambda_{j_\mathrm{S}}\right)^{n-1} \trc_\mathrm{B}\left( \prod_{s=1}^{n} \trc_\mathrm{A} \Ket{\lambda_{i_\mathrm{S}}}\Bra{\lambda_{j_\mathrm{S}}} \right) = 0\,. \\
\end{eqnarray}
\end{subequations}
\end{widetext}
Condition in \eqref{eq:Cond_Normalization} is essentially the preservation of the normalization condition at any time $t$, which is trivial because $c_i$'s are time-independent.
Condition in \eqref{eq:Cond_Purity} expresses the conservation of the linear entropy at every time. Let us  analyze this condition more carefully.
Note that, due to the linear independence of the exponential functions we can simplify condition \eqref{eq:Cond_Purity} as follows:
let us call two sets of indices $\{i_1,j_1,i_2,j_2\}$ and $\{i'_1,j'_1,i'_2,j'_2\}$ equivalent iff
$$   \lambda_{i_1}-\lambda_{j_1}+\lambda_{i_2}-\lambda_{j_2} = \lambda_{i'_1}-\lambda_{j'_1}+\lambda_{i'_2}-\lambda_{j'_2} , $$
and denote the class  indices equivalent to $\{i_1,j_1,i_2,j_2\}$ by $[i_1,j_1,i_2,j_2]$. Now, \eqref{eq:Cond_Purity} implies the following condition: for any $\{\underline{i_1}, \underline{j_1}, \underline{i_2}, \underline{j_2}\}$ such that $\lambda_{i_1}+\lambda_{i_2} \neq \lambda_{j_1}+\lambda_{j_2}$ one has:
\begin{widetext}
\begin{equation}\label{}
  \sum_{\{i_1,j_1,i_2,j_2\} \in [\underline{i_1}, \underline{j_1}, \underline{i_2}, \underline{j_2}]} c_{i_1} c_{j_1}^*c_{i_2} c_{j_2}^* \,\trc_\mathrm{B}\left(  \trc_\mathrm{A} \Ket{\lambda_{i_1}}\Bra{\lambda_{j_1}} \trc_\mathrm{A} \Ket{\lambda_{i_2}}\Bra{\lambda_{j_2}} \right) = 0 \ .
\end{equation}
\end{widetext}
It is easy to verify that when $H$ does not contain any interaction term, then these conditions are automatically satisfied. Indeed, if $H=H_\mathrm{A}+H_\mathrm{B}$ then there exists a set of eigenvectors which are nothing but products of states of $\mathcal{H}_\mathrm{A}$ and $\mathcal{H}_\mathrm{B}$: $\Ket{\lambda_k}= \Ket{\phi_k}_\mathrm{A}\otimes \Ket{\psi_k}_\mathrm{B}$, which implies
\begin{equation}
\trc_\mathrm{B}\left( \trc_\mathrm{A} \Ket{\lambda_{i_1}}\Bra{\lambda_{j_1}} \trc_\mathrm{A} \Ket{\lambda_{i_2}}\Bra{\lambda_{j_2}}\right) = \delta_{i_1 j_2}\delta_{i_2 j_1}\, ,
\end{equation}
and hence either $\trc_\mathrm{B}\left( \trc_\mathrm{A} \Ket{\lambda_{i_1}}\Bra{\lambda_{j_1}} \trc_\mathrm{A} \Ket{\lambda_{i_2}}\Bra{\lambda_{j_2}}\right) =0$ or $\lambda_{i_1}+\lambda_{i_2}  = \lambda_{j_1}+\lambda_{j_2}$, and this ensures that \eqref{eq:Cond_Purity} is satisfied.

For the generic $k$ one gets: for any $\{i_1, j_1, \ldots , i_k, j_k\}$ such that $\lambda_{i_1}+ \ldots +\lambda_{i_k} \neq \lambda_{j_1}+ \ldots + \lambda_{j_k}$ one has
\begin{widetext}
\begin{equation}\label{}
   \sum_{\{i_1,j_1,\ldots,i_k,j_k\} \in [\underline{i_1},\underline{j_1},\ldots,\underline{i_k},\underline{j_k}]}c_{i_1} c_{j_1}^*c_{i_2} c_{j_2}^*... c_{i_k} c_{j_k}^*  \trc_\mathrm{B}\left( \trc_\mathrm{A} \Ket{\lambda_{i_1}}\Bra{\lambda_{j_1}} \trc_\mathrm{A} \Ket{\lambda_{i_2}}\Bra{\lambda_{j_2}} ... \trc_\mathrm{A} \Ket{\lambda_{i_k}}\Bra{\lambda_{j_k}}  \right) = 0\ ,
\end{equation}
\end{widetext}
where now the equivalence of indexes $\{i_1,j_1,\ldots,i_k,j_k\}$ and $\{i'_1,j'_1,\ldots,i'_k,j'_k\}$ is defined by
$$  \sum_{l=1}^k (\lambda_{i_l}- \lambda_{j_l}) = \sum_{l=1}^k (\lambda_{i'_l}- \lambda_{j'_l}) . $$
Again, one immediately verifies that when $H$ does not contain any interaction term, then these conditions are automatically satisfied.

It is also the case to point out that all such conditions, for all values of $k$, are automatically satisfied if all $c_i$'s are zero but one: $c_i = \delta_{ip}$ for a given $p$. This corresponds to the trivial result that all the eigenstates of the Hamiltonian are GIFE states.

{\bf An Example} --- In order to illustrate our strategy for
finding the GIFE states of a given Hamiltonian, let us consider
the very simple example of two interacting two-level systems
described by the following Hamiltonian:
\begin{eqnarray}
H = \omega_\mathrm{A} \sigma_z^{(\mathrm{A})} + \omega_\mathrm{B} \sigma_z^{(\mathrm{B})} + \gamma (\sigma_-^{(\mathrm{A})}\sigma_+^{(\mathrm{B})}+\sigma_-^{(\mathrm{A})}\sigma_+^{(\mathrm{B})})\, .
\end{eqnarray}
One finds for the eigenvalues
\begin{eqnarray*}
  \lambda_1 &=& \omega_\mathrm{A}+\omega_\mathrm{B} , \\
  \lambda_2 &=& - \omega_\mathrm{A}- \omega_\mathrm{B} \\
  \lambda_3 &=& -\sqrt{\gamma^2+(\omega_\mathrm{B}-\omega_\mathrm{A})^2} \\
  \lambda_4 &=& \sqrt{\gamma^2+(\omega_\mathrm{B}-\omega_\mathrm{A})^2}
\end{eqnarray*}
together with the corresponding  eigenvectors
\begin{eqnarray}
\nonumber
\Ket{\lambda_1}  &=&  \Ket{+}_\mathrm{A} \Ket{+}_\mathrm{B}   \\
\nonumber
\Ket{\lambda_2}  &=&  \Ket{-}_\mathrm{A} \Ket{-}_\mathrm{B}      \\
\nonumber
\Ket{\lambda_3}  &=&  N_3[(\omega_\mathrm{B}-\omega_\mathrm{A} - \sqrt{\gamma^2+(\omega_\mathrm{B}-\omega_\mathrm{A})^2})\Ket{-}_\mathrm{A} \Ket{+}_\mathrm{B}  \\ \nonumber
                    &+& \gamma \Ket{+}_\mathrm{A}\Ket{-}_\mathrm{B}] ,   \\
\nonumber
\Ket{\lambda_4}  &=&   N_4[(\omega_\mathrm{B}-\omega_\mathrm{A} + \sqrt{\gamma^2+(\omega_\mathrm{B}-\omega_\mathrm{A})^2})\Ket{-}_\mathrm{A} \Ket{+}_\mathrm{B}  \\
\nonumber
     &+& \gamma \Ket{+}_\mathrm{A} \Ket{-}_\mathrm{B}]\, ,
\end{eqnarray}
with $N_3$ and $N_4$ being suitable normalization factors. It is now straightforward to obtain conditions for coefficients $c_k$ in $\Ket{\chi}=\sum_k c_k \Ket{\lambda_k}$, which guarantee the preservation of linear entropy and hence provide GIFE state:
\begin{subequations}
\begin{eqnarray}
   \label{eq:Example_Proper_GIFE}
   && c_1 = c_3=0\, , \qquad  \mbox{and arbitrary} \ \  c_2, c_4\,;  \\
   && c_2 = c_3=0\, ,  \qquad \mbox{and arbitrary}\ \  c_1, c_4\,;  \\
   \label{eq:Example_IFE}
   && c_3 = c_4=0\, , \qquad  \mbox{and arbitrary} \ \ c_1, c_2\,;  \\
   && c_1 = c_4=0\, , \qquad  \mbox{and arbitrary} \ \ c_2, c_3\,;  \\
   && c_2 = c_4=0\, , \qquad  \mbox{and arbitrary} \ \ c_1, c_3\,.
\end{eqnarray}
\end{subequations}

These solutions show in a clear way what we have anticipated, that
all the eigenstates of the Hamiltonian are GIFE states. Some of
the conditions we have found for the coefficients give IFE states:
for example \eqref{eq:Example_IFE} gives rise to IFE states, as it
is quite easy to see. On the contrary, other conditions, like for
example \eqref{eq:Example_Proper_GIFE}, give rise to GIFE, since
the state $c_2 \e^{-\i \lambda_2 t} \Ket{\lambda_2}+c_4 \e^{-\i
\lambda_4 t} \Ket{\lambda_4}$ can never be considered as
essentially evolving according to the free Hamiltonian of the
system, unless $\omega_\mathrm{A}=\omega_\mathrm{B}$. Indeed, for
example, the complete evolution shows that the states
$\Ket{-}_\mathrm{A}\Ket{+}_\mathrm{B}$ and
$\Ket{+}_\mathrm{A}\Ket{-}_\mathrm{B}$ accumulate the same phase,
while the free evolution alone would give to them different
phases, when $\omega_\mathrm{A}\not=\omega_\mathrm{B}$. An example
of two possible effective Hamiltonians is given by:
$H_\mathrm{A}^{\mathrm{eff}}=\tilde{\omega}
\sigma_z^{(\mathrm{A})}$ and $H_\mathrm{B}^{\mathrm{eff}}=
\tilde{\omega} \sigma_z^{(\mathrm{B})}$, with
$\tilde{\omega}=(\lambda_4-\lambda_2)/2$.

\section{Conclusions}\label{sec:Discussion}

In this paper we have explored the connection between the two
concepts of Interaction-Free Evolutions and Decoherence-Free
Subspaces, bringing to the light similarities and differences.
The very first difference between
IFE and DFS is given by the context and the class of systems they
refer to, in the sense that talking about DFS requires that one of the two subsystems is the environment and that the dynamics of the small system is unitary for all possible states of the environment; such restriction do not apply to  IFE.
Therefore, when the system is made of a small system and its
environment, both IFE and DFS are in principle possible. Since one
could think that the existence of one of the two class of states
implies the existence of the other one, we have explored this
possible connection, pointing out some general
statements. We have brought
to the light the fact that, though the two concepts are both
related to the idea that somehow the interaction between the two
subsystems is not felt, the two concepts are quite independent. In
fact, it can happen that DFS are present but no IFE, that IFE are
present but no DFS, and that both IFE and DFS are present. This
independence has been discussed in sec.~\ref{sec:ife_vs_dfs}.

The dynamics of a system prepared in a DFS is essentially governed
by the free Hamiltonian of the system, as if the interaction were
not present, but this notion has been generalized in the
literature including the case where the system undergoes a unitary
evolution even if such an evolution is generated by an effective
Hamiltonian that differs from the free one (in this cases we talk
about generalized DFS, i.e., GDFS). On this basis, we have
extended the concept of IFE introducing the idea of Generalized
Interaction-Free Evolutions (GIFE) and provided a characterization of such class of evolutions. In particular, starting from noting that in such evolutions the amount of entanglement between the two subsystems must be preserved, we have found that a set of functionals can be considered that are necessarily preserved during any kind (properly generalized or not) of interaction-free evolution. On this basis, we have developed a strategy for systematically obtain all possible GIFE states for any given Hamiltonian. Then we have applied such strategy on a specific (simple) situation.

\section*{Acknowledgements}

We thank anonymous referee for valuable comments. DC was partially supported by the National Science Center project
DEC-2011/03/B/ST2/00136.

\end{document}